\documentstyle[twocolumn,epsf,aps,prb]{revtex}
\begin{document}
\draft
\wideabs{
\title{
Na-site substitution effects on the 
thermoelectric properties of NaCo$_2$O$_4$
}

\author{
T. Kawata,$^1$ Y. Iguchi,$^1$ T. Itoh,$^1$ K. Takahata,$^1$
and I. Terasaki$^{1,2,}$\cite{byline}
}

\address{
$^1$Department of Applied Physics, 
Waseda University, Tokyo 169-8555, Japan\\
$^2$Precursory Research for Embryonic Science and
Technology, Japan Science Technology, Tokyo 102-0074, Japan
}

\date{\today}

\maketitle

\begin{abstract}
The resistivity and thermopower of Na$_{1+x}$Co$_2$O$_4$ and
Na$_{1.1-x}$Ca$_x$Co$_2$O$_4$ are measured and analyzed.
In Na$_{1+x}$Co$_2$O$_4$, whereas the resistivity increases with $x$, 
the thermopower is nearly independent of $x$.
This suggests that the excess Na is unlikely to supply carriers, and 
decreases effective conduction paths in the sample.
In Na$_{1.1-x}$Ca$_x$Co$_2$O$_4$, the resistivity and the thermopower
increase with $x$, 
and the Ca$^{2+}$ substitution for Na$^+$ reduces
the majority carriers in NaCo$_2$O$_4$.
This means that they are holes,
which is consistent with the positive sign of the thermopower.
Strong correlation in this compound is evidenced by 
the peculiar temperature dependence of the resistivity.
\end{abstract}


} 


There appears a growing interest to a hunt 
for new thermoelectric (TE) materials, \cite{Mahan}
reflecting urgent needs for a new energy-conversion system 
in harmony with our environments.
A TE material generates electric power
in the presence of temperature gradient
through the Seebeck effect,
and pumps heat in the presence of electric current through the 
Peltier effect. 
A serious drawback is the low conversion efficiency:
It is characterized by the so-called ``figure of merit'' 
$Z=S^2/\rho\kappa$, where $S$, $\rho$ and $\kappa$ are
the thermopower, resistivity and thermal conductivity of a TE material,
respectively.
In other words, a good TE material is a material that shows
large $S$, low $\rho$ and low $\kappa$. 
However, a high value of $Z$ is difficult to realize, because
the three parameters cannot be changed independently. 
To overcome this difficulty, 
a number of new concepts and new materials have been examined.

Recently we have observed that a layered cobalt oxide 
NaCo$_2$O$_4$ exhibits unusually large $S$ 
(100 $\mu$V/K at 300 K) accompanied 
by low $\rho$ (200 $\mu\Omega$cm at 300 K)
along the direction parallel to the CoO$_2$ plane. \cite{Terra}
NaCo$_2$O$_4$ belongs to a layered Na bronze Na$_x$CoO$_2$,
which was studied as a cathode
for sodium batteries. \cite{Delmas}
During the characterization,
Molenda {\it et al.} \cite{Molenda} first found
a large $S$ in Na$_{0.7}$CoO$_2$.
Although they noticed that $S$ was anomalously large,
they did not mention a possibility for a TE material.
Their samples were polycrystals, the resistivity of which was
2-4 m$\Omega$cm at 300 K, much higher than that of our crystals.
Our finding is that the carrier density ($n$)
is of the order of $10^{21}-10^{22}$ cm$^{-3}$,
and is two orders of magnitude larger than $n$ of 
conventional TE materials.
This is difficult to understand in the framework of 
a conventional one-electron picture, 
and may indicate a way to 
get a good TE material other than the conventional approach.
We have proposed that strong electron-electron correlation 
plays an important role in the enhancement of the thermopower
of NaCo$_2$O$_4$.

Even in a correlated system, 
we can expect that a conductor of low $n$ will have a large $S$,
because the diffusive part of $S$ is the transport entropy,
of the order of $k_BT/E_F$, where $E_F$ 
is the Fermi energy. \cite{Uher}
Thus it would be tempting to improve 
the TE properties in NaCo$_2$O$_4$ by decreasing $n$.
We easily think of three ways to change $n$ in NaCo$_2$O$_4$,
{\it i.e.}, (i) doping of excess Na$^+$, 
(ii) the substitution of Ca$^{2+}$ for Na$^+$,
and (iii) the change of the oxygen content.
Among them, we will discard the idea of (iii), 
because it will seriously deteriorate the conduction paths 
consisting of Co and O.
Here we report on the resistivity and thermopower of
Na$_{1+x}$Co$_2$O$_4$ and Na$_{1.1-x}$Ca$_x$Co$_2$O$_4$ 
to study the doping effects.


We prepared 
polycrystalline samples  of Na$_{1+x}$Co$_2$O$_4$ and
Na$_{1.1-x}$Ca$_x$Co$_2$O$_4$ by solid state reaction.
Since Na is volatile, we added 10 \% excess Na.
Namely we expected the starting composition of Na$_{1.1}$Co$_2$O$_4$ 
to be NaCo$_2$O$_4$.
An appropriate mixture of Na$_2$CO$_3$, CaCO$_3$, Co$_3$O$_4$
was thoroughly ground, sintered at 860--920$^{\circ}$C for 12 h in air.
The sintered powder was then pressed into a pellet,
and sintered again at 800--920$^{\circ}$C for 6 h in air.

The x-ray diffraction (XRD) was measured 
using a standard diffractometer with Fe K$_{\alpha}$ radiation 
as an x-ray source in the $\theta -2\theta$ scan mode.
Note that Cu K$_{\alpha}$ radiation is not suitable for
this compound, because it emits the fluorescent x-ray of 
Co to make a high noise in the XRD pattern.
$\rho$ was measured through a four-probe method,
in which the electric contacts with a contact resistance of 1 $\Omega$ 
were made with silver paint (Dupont 4922).
$S$ was measured using a steady-state technique.
Temperature gradient ($\sim$0.5 K/cm) was generated 
by a small resistive heater pasted on one edge of the sample,
and was monitored by a differential thermocouple 
made of copper-constantan.
A thermopower of voltage leads was carefully subtracted.
Temperature ($T$) was controlled from 4.2 to 300 K in a liquid He 
cryostat, and was monitored with a CERNOX resistance thermometer.


Figure 1 shows typical XRD patterns
of the prepared samples.
Almost all the peaks are indexed as the P2 phase
reported by Jansen and Hoppe,\cite{JH}
though a tiny trace of impurity phases is 
detected as marked with $*$ in Fig. 1.
Note that all the XRD patterns are nearly the same,
which means that XRD is not very powerful
for the sample characterization.
Thus the best way to characterize the samples is 
to measure their thermoelectric properties directly.
Usually an impurity phase including Na will be Na$_2$O,
and exist as deliquesced NaOH (Na$_2$O +H$_2$O).
We think, however, that Na$_2$O is not a major impurity phase 
for the present case.
The samples are stable enough to handle in air, and
the contact resistance and the surface do not deteriorate
against several-hour exposure to the air.

Figure 2(a) shows $\rho$ for  Na$_{1+x}$Co$_2$O$_4$ 
plotted as a function of $T$.
Both the magnitude and the $T$ dependence are consistent
with previous studies. \cite{Molenda,Tanaka}
All the samples show a metallic conduction down to 4.2 K
without any upturn at low temperatures.
This suggests that the conduction paths are not 
disturbed by the doped excess Na.
The $T$ dependence of $\rho$ roughly resembles 
the in-plane resistivity for single-crystal NaCo$_2$O$_4$, \cite{Terra}
implying that the conduction of polycrystals is
mainly determined by the in-plane conduction.
Note that $\rho$ for $x$=0 is higher than $\rho$ for $x$=0.1,
which suggests that a small amount of Na is evaporated
through the sintering process.

Contrary to the change of $\rho$ with $x$, 
$S$ for Na$_{1+x}$Co$_2$O$_4$ is
nearly independent of $x$ as shown in Fig. 2(b).
This indicates that $n$ remains intact by doping Na.
It is, at first sight, unusual
why the doped monovalent Na$^+$ does not change $n$.
We point out two possibilities:
One is that the excess Na is excluded from the crystal
to increase the resistance at the grain boundary,
and the other is that it is in the grain
to make an insulating phase nearby.
Note that  NaCoO$_2$ (corresponding to $x$=1) 
is an insulator.\cite{Delmas}
In both cases, excess Na cations decrease the number of conduction paths to
reduce the effective cross section for the current.

Making a remarkable contrast to Fig. 2(a),
Figure 3 (a) shows a drastic change of $\rho$ for
Na$_{1.1-x}$Ca$_x$Co$_2$O$_4$ with $x$.
Above 50 K, while $\rho$ for $x$=0  shows a positive curvature, 
$\rho$ for $x$=0.35 shows a negative curvature to saturate near 300 K.
Unlike the case of the excess Na,
the residual resistivity, though not well-defined,
tends to increase with $x$,
which means that Ca acts as a scattering center.
$S$ is also increased with $x$ as shown in Fig. 3(b).
Considering that both $\rho$ and $S$ increase with Ca,
we conclude that the substitution of Ca$^{2+}$ for Na$^+$
decreases the carriers.
Namely the majority carrier of NaCo$_2$O$_4$ is a hole,
which is consistent with the transport properties of
Na$_{0.7}$CoO$_{2-\delta}$. \cite{Molenda}
As expected, the TE properties are (slightly) improved by decreasing $n$,
and $S^2/\rho$ is maximized at $x$=0.15.

One may notice that Na$_{1.1}$Co$_2$O$_4$ shows different $\rho$
between Figs. 2 and 3.
The magnitude of $\rho$ was scattered from batch to batch,
possibly because the control of the grain growth is difficult.
(Thermopower is a quantity less affected by grain boundaries, and
the measured $S$ was independent of batches within experimental errors.)
To see the reproducibility we made Na$_{1.1}$Co$_2$O$_4$ as a reference
at every preparation run.
Figure 4 shows $\rho$ for Na$_{1.1}$Co$_2$O$_4$ prepared in different runs,
where the magnitude of $\rho$ is scattered beyond experimental errors ($\sim$10\%).
We note that the relative change of $\rho$ 
among the same batch is reproducible,
and the $T$ dependence is essentially identical from batch to batch.
All the $\rho -T$ data in Fig. 4 normalized at 295 K
fall into a single curve, as shown in the inset of Fig. 4.

Figure 5 shows $\rho$ of Na$_{1.1}$Co$_2$O$_4$ in Fig. 2(a)
is plotted in a log-log scale.
Since $\rho$ is linear below 50 K and above 80 K,
$\rho$ is proportional to $T^p$ in the two regions.
From fitting $\rho$ by $T^p$, we estimated $p$ 
to be 0.67 below 50 K and 1.2 above 80 K
(see the solid and dashed lines in Fig. 5).
We will remark three points on the $T$ dependence of $\rho$.
First, it is a piece of evidence for strong correlation that 
$\rho$ continues to decrease with decreasing $T$ down to 4.2 K
where no phonons are thermally excited.
At least we can say that the conduction 
in this system is not dominated by the 
conventional electron-phonon scattering.
Secondly the $T$ dependence of $\rho$ of this system
is not typical for strongly correlated systems.
In usual strongly correlated systems,
resistivity and electron-electron scattering 
are proportional to $(k_BT/E_F)^2$. 
Most of heavy fermions, \cite{KW} 
organic conductors, \cite{Dressel} transition-metal oxides \cite{TM,TM2}
show $\rho \propto T^2$.
As shown in the inset of Fig. 5, 
$\rho$ for Na$_{1.1}$Co$_2$O$_4$ is not proportional to $T^2$ 
at any temperatures.
A prime exception is the $T$-linear resistivity in high-$T_c$
superconductors. \cite{HTSC}
Actually, $\rho$ and $S$ of NaCo$_2$O$_4$ are qualitatively consistent with
some theories for high-$T_c$ superconductors. \cite{Moriya,Miyake}
In particular, $\rho$ of Na$_{1.1-x}$Ca$_x$Co$_2$O$_4$ 
can be explained by adjusting the parameters in Ref. \onlinecite{Moriya}. 
Thirdly all the samples show no indication of localization.
This means that the mean free path (MFP) of the carriers is much longer than 
the lattice parameters, \cite{Terra} and that the carriers do not feel 
the disorder in the Na layer.
On the other hand, phonons will be affected
by the disorder in the Na layer, 
since the disorderd Na$^+$ ions make ionic bonding with 
adjacent O$^{2-}$ ions.
In fact, a preliminary measurement has revealed  that $\kappa$ for
Na$_{1.1}$Co$_2$O$_4$ is as low as 10 mW/cmK, \cite{Yakabe}
suggesting that MFP of the phonons is of the order of 
the lattice spacing. 
Thus MFP of the carriers is much longer
than MFP of the phonons in NaCo$_2$O$_4$.
We therefore propose that this material is a new class of 
``electron crystals and phonon glasses''. \cite{Slack}

Finally let us comment on strong correlation.
Since the diffusive part of $S$ corresponds to the 
transport entropy, as mentioned above, 
larger electronic specific heat can give larger $S$. 
Thus $S$ would be enhanced if the carriers
could couple with some outside entropy such as 
optical phonon, spin fluctuation, or orbital fluctuation.
Recently a similar scenario is independently proposed 
by Palsson and Kotliar. \cite{Palsson}
Heavy fermions or valence-fluctuation systems are indeed the case,
some of which show large $S$. \cite{Mahan}
Very recently Ando {\it et al.} \cite{Ando} 
have measured the specific heat of
Na$_{1.1-x}$Ca$_x$Co$_2$O$_4$ at low temperatures,
and have found a large electronic specific heat of 48 mJ/mol~K$^2$,
which is one order of magnitude larger than conventional metals.


In summary, 
we have prepared polycrystals of 
Na$_{1+x}$Co$_2$O$_4$ and Na$_{1.1-x}$Ca$_x$Co$_2$O$_4$,
and measured the resistivity and thermopower from 4.2 to 300 K.
The excess Na and the substituted Ca affect the transport properties
of NaCo$_2$O$_4$ differently.
The former seems to decrease the effective conducting region,
and the latter decreases the carrier density.
The temperature dependence of the resistivity is drastically
changed by substituting Ca, which strongly suggests
that the scattering mechanism depends on the carrier density.
Combining this with the peculiar temperature dependence of the
resistivity, we conclude that strong electron-electron correlation
plays an important role in this compound.


The authors would like to thank
Y. Ando, K. Segawa and N. Miyamoto
for collaboration.
They are also indebted to 
H. Yakabe, K. Fukuda, K. Kohn, S. Kurihara, S. Saito,
and M. Takano for the fruitful discussion.



\begin{figure}
\centerline{\epsfxsize=7cm 
\epsfbox{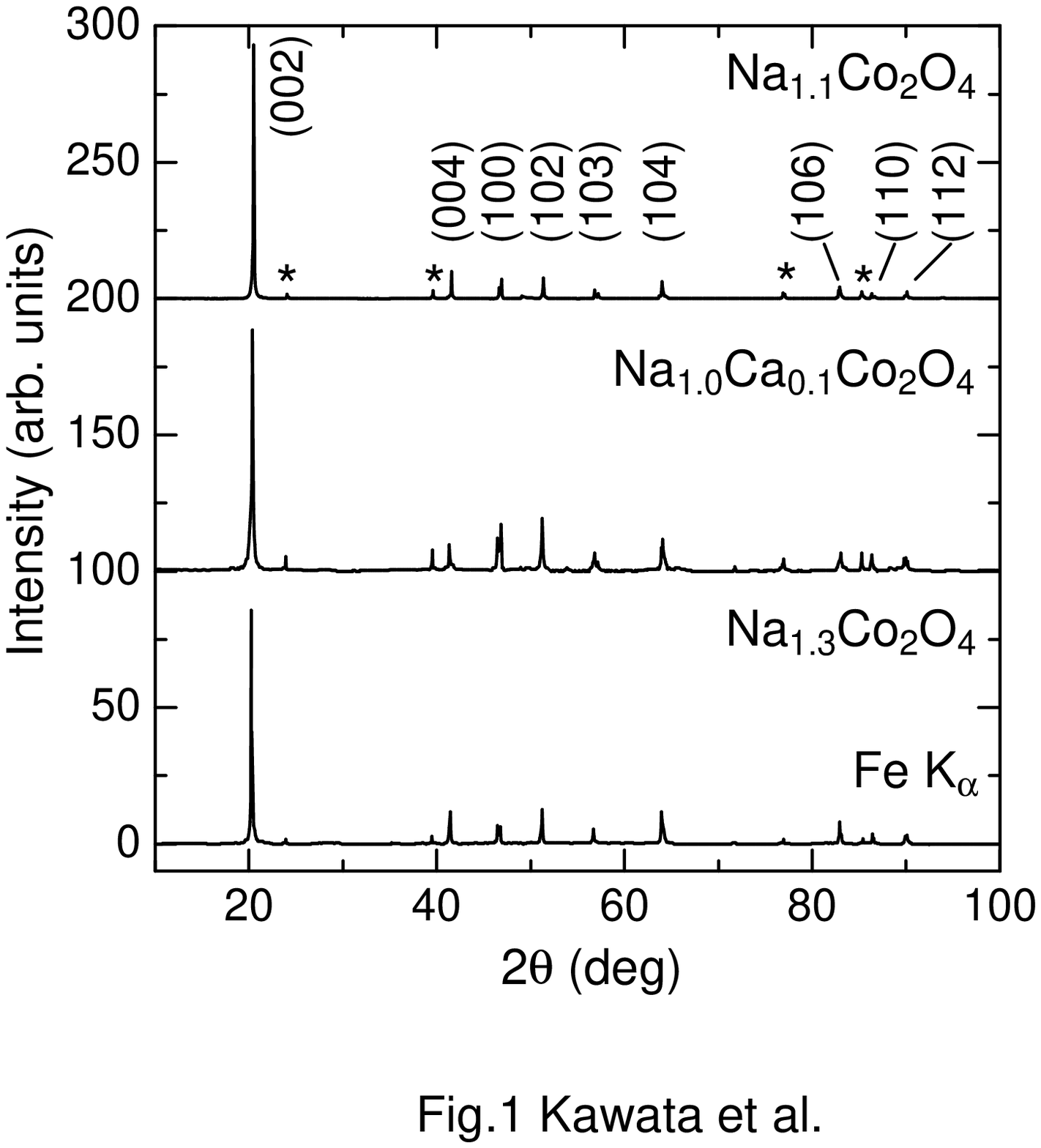}
}
\caption{
The  x-ray diffraction patters of 
Na$_{1.1}$Co$_2$O$_4$,
Na$_{1.3}$Ca$_x$Co$_2$O$_4$ and
Na$_{1.0}$Ca$_{0.1}$Co$_2$O$_4$.
The Fe K$_{\alpha}$ radiation is used as an x-ray source.
Most of the peaks are indexed as the P2 phase,
while a few peaks of impurity phases are marked with $*$. 
}
\end{figure}

\begin{figure}
\centerline{\epsfxsize=7cm 
\epsfbox{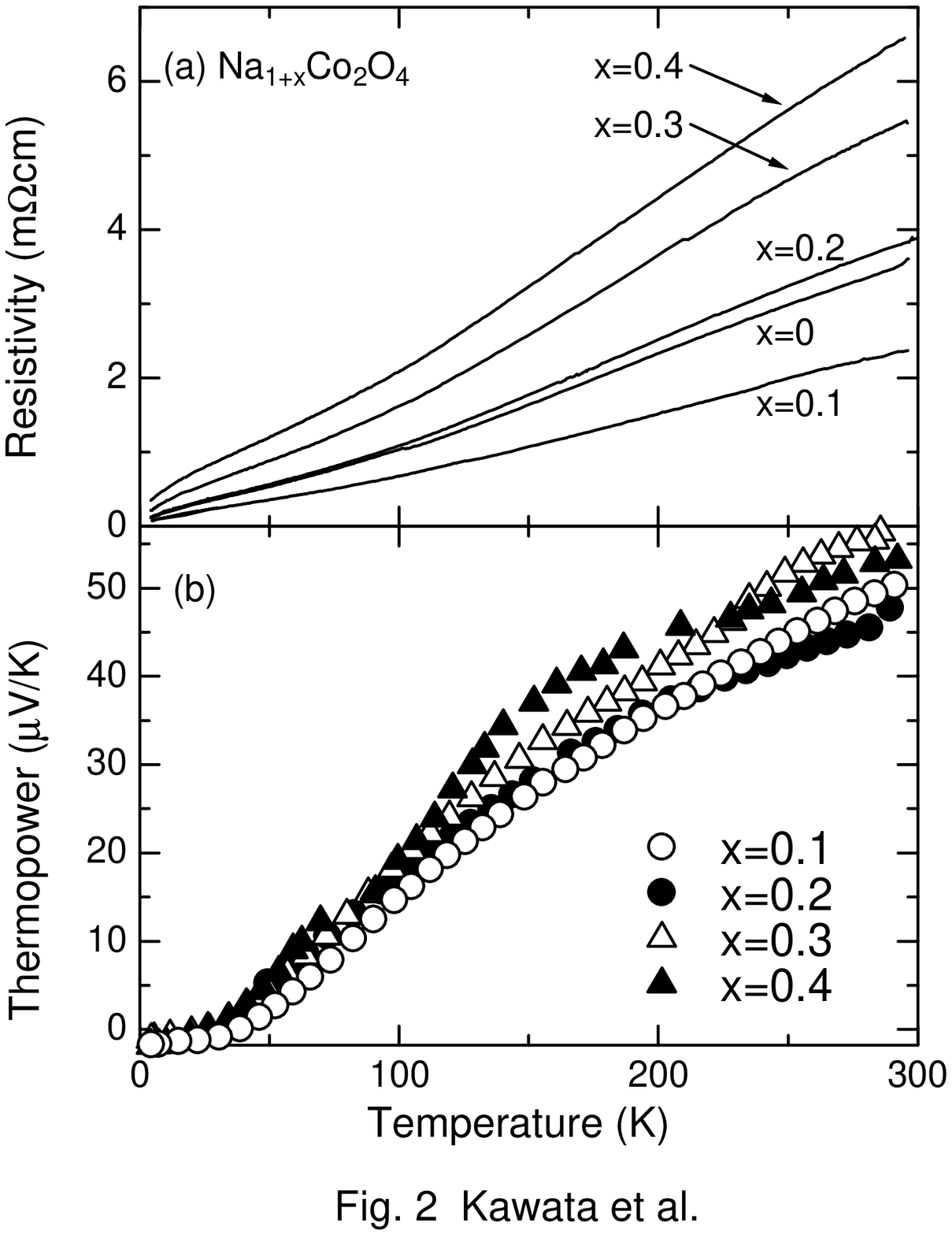}
}
\caption{
(a) Resistivity 
and (b) Thermopower of Na$_{1+x}$Co$_2$O$_4$
plotted as a function of temperature.
}
\end{figure}

\begin{figure}
\centerline{\epsfxsize=7cm 
\epsfbox{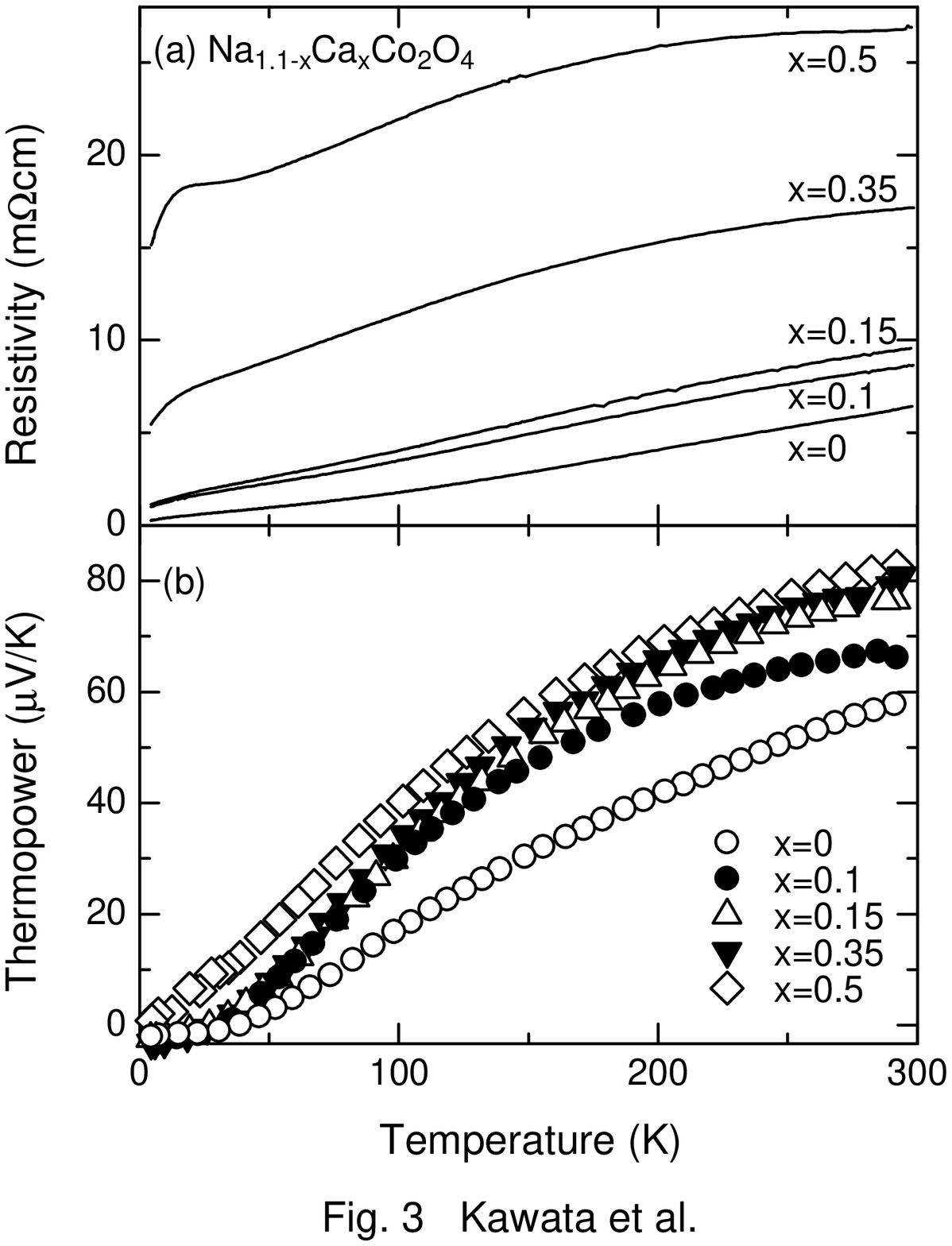}
}
\caption{
(a) Resistivity 
and (b) Thermopower of Na$_{1.1-x}$Ca$_x$Co$_2$O$_4$
plotted as a function of temperature.
}
\end{figure}

\begin{figure}
\centerline{\epsfxsize=7cm 
\epsfbox{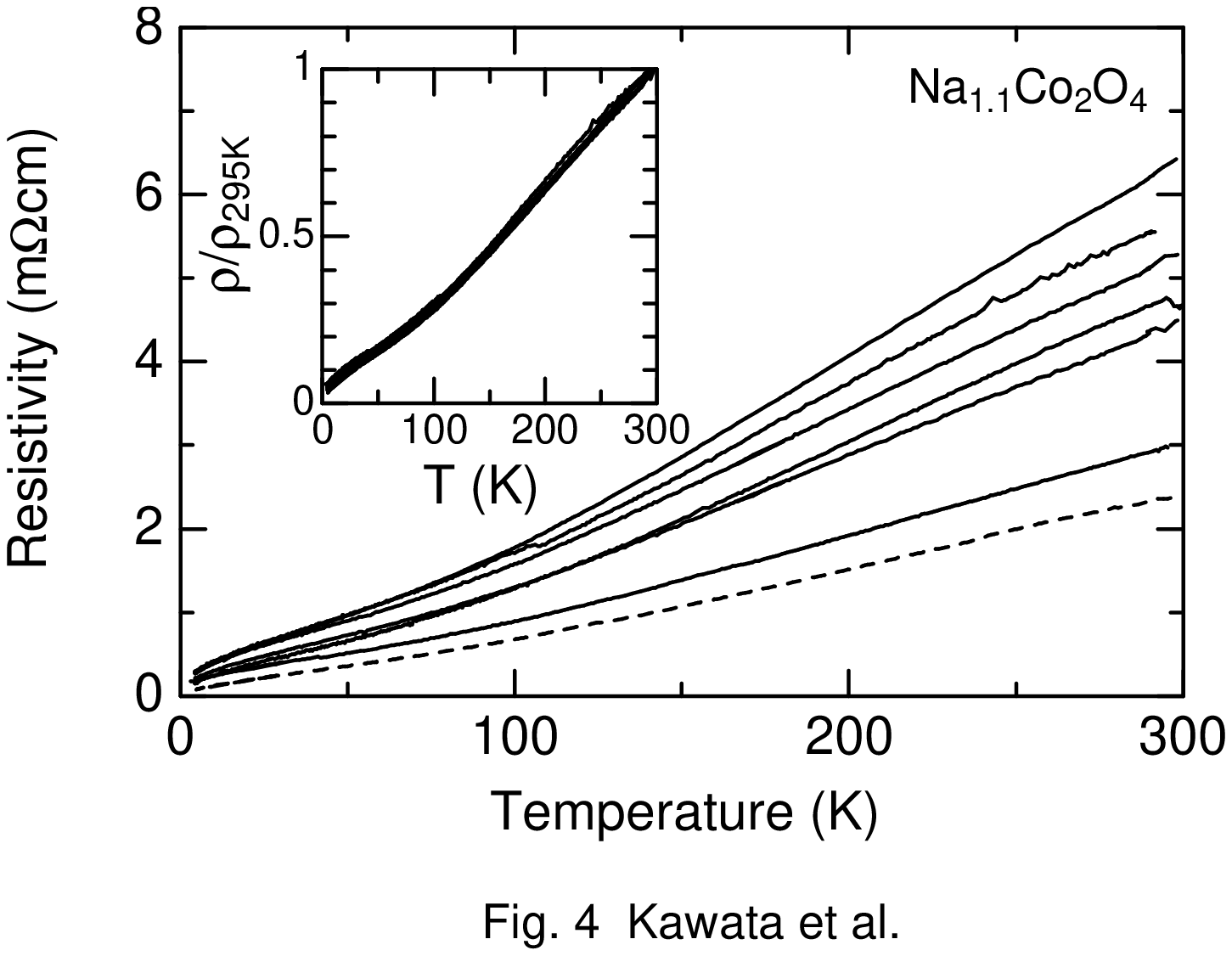}
}
\caption{
Resistivity of Na$_{1.1}$Co$_2$O$_4$ prepared in different batches.
The solid curves represent samples sintered at 860$^{\circ}$C,
and the dashed curve represents a sample sintered at 920$^{\circ}$C.
The magnitude of resistivity is scattered beyond 
experimental errors ($\sim$ 10\%).
The resistivity normalized at 295 K is shown in the inset.
}
\end{figure}

\begin{figure}
\centerline{\epsfxsize=7cm 
\epsfbox{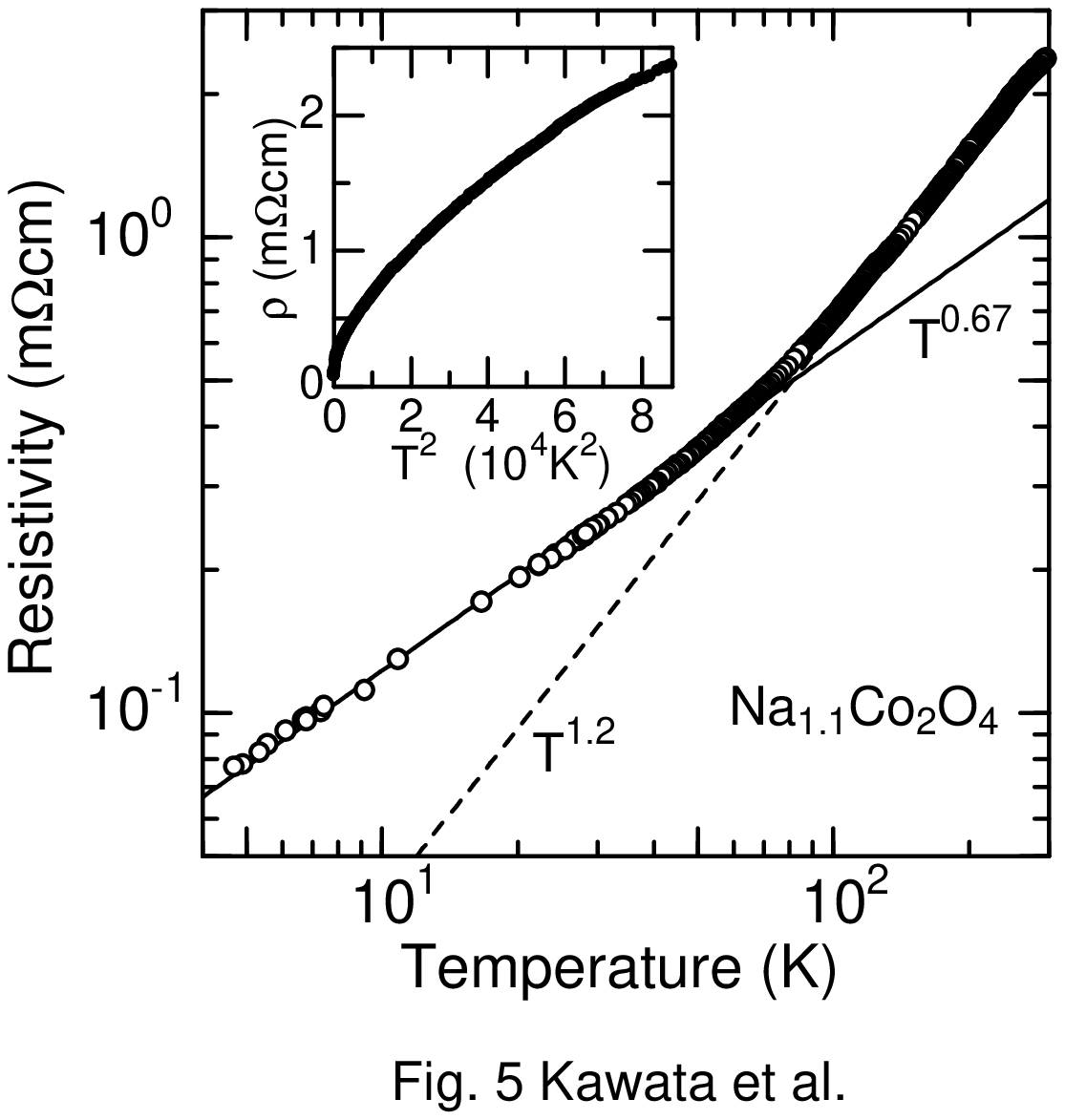}
}
\caption
{Log-log plot of the resistivity of Na$_{1.1}$Co$_2$O$_4$.
The data are the same as in Fig. 2(a).
The solid and dashed lines represent
$\rho\propto T^{0.67}$ and $\rho\propto T^{1.2}$, respectively.
The inset shows $\rho$ plotted as a function of $T^2$. 
}
\end{figure}

\end{document}